\documentclass[pra,superscriptaddress,showpacs,twocolumn]{revtex4}
\usepackage{amssymb}

\def\>{\ensuremath{\rangle}}
\def\<{\ensuremath{\langle}}

\def\e{\ensuremath{\mathcal{E}}}

\def\h{\ensuremath{\mathcal{H}}}

\def\tr{{\rm Tr}}

\begin{document}
\title{Optimal dense coding with arbitrary pure entangled states}
\author{Yuan Feng}
\email{feng-y@tsinghua.edu.cn}
\author{Runyao Duan}
\email{dry02@mails.tsinghua.edu.cn}
\author{Zhengfeng Ji}
\email{jizhengfeng98@mails.tsinghua.edu.cn}

\affiliation{State Key Laboratory of Intelligent Technology and Systems,\\
\small Department of Computer Science and Technology,\\
\small Tsinghua University, Beijing, China, 100084}

\begin{abstract}
We examine dense coding with an arbitrary pure entangled state
sharing between the sender and the receiver. Upper bounds on the
average success probability in approximate dense coding and on the
probability of conclusive results in unambiguous dense coding are
derived. We also construct the optimal protocol which saturates the
upper bound in each case.
\end{abstract}

\maketitle

\section{Introduction}
Dense coding \cite{BW92} is a communication protocol which, making
use of entanglement shared in prior between the sender (Alice) and
the receiver (Bob), can improve the classical capacity of a
noiseless quantum channel. In the original protocol proposed by
Bennett and Wiesner in Ref.\cite{BW92}, with the assistance of a
maximally entangled qubit pair, Alice can send faithfully 2 bits of
classical information to Bob by sending a single qubit. Notice that
it is only possible for Alice to send 1 bit of classical information
by sending a qubit without the assistance of entanglement
\cite{Hol73,YO93}. The use of entanglement in this protocol doubles
the classical capacity of the noiseless quantum channel. In the same
paper, the authors generalized the protocol to transmit faithfully
$2\log_2d$ bits of classical information, making use of a maximally
entangled state in $d$-level quantum system.

The original dense coding protocol has been generalized by other
authors in various directions, such as the case of continuous
variables \cite{BK00,ZXP02} and multipartite communication
\cite{HLG01,LLTL02,BAL+04}. Maximally entangled states are very
difficult to prepare and store in practical applications, so it is
important to consider the performance of dense coding when the
states shared between the sender and the receiver are partially
entangled. On the other hand, when only a partially entangled state
is available, it is easy to check that perfect dense coding
presented in Ref.\cite{BW92} is impossible. That is, we can not hope
to transmit faithfully $2\log_2d$ bits of classical information with
certainty, provided that a partial entanglement in $d$-level quantum
system is shared. To deal with the problem of dense coding with
arbitrary pure entangled states, Hao et al. \cite{HLG00} initialized
the exploration of probabilistic dense coding for qubit case, in
which the protocol succeeds only with some probability less than 1.
Probabilistic dense coding was further extended to higher
dimensional case by Pati et al. \cite{PPA05} and Wu et al.
\cite{WCSG06}. Another clue of research, discussed in
Refs.\cite{MOR05} and \cite{JFDY06}, pays attention to investigation
of the relation between the form and the amount of shared
entanglement and the maximal size of alphabet which can be
faithfully transmitted from Alice to Bob. Rather surprisingly,
approximate dense coding in which there exists some probability of
error has not considered in the literature.

In this paper, we consider dense coding with an arbitrary pure
entangled state in both approximate case and unambiguous case. Our
contribution is twofold: First, we derive an upper bound on the
average success probability of dense coding in approximate case. An
explicit protocol which saturates this bound is also given. Second,
we consider the case of unambiguous dense coding and derive the
optimal conclusive probability which turns out to be a constant for
any input signal. We also construct explicitly a protocol in which
this optimal probability is achieved.

\section{Strategies of imperfect dense coding}

This section devotes to the clarification of the differences between
imperfect dense coding strategies presented in the literature and
those proposed in this paper. To make the statements more rigorous,
we first formulate the problem of dense coding as follows.

Suppose Alice and Bob share in prior an entangled pure state
$|\Phi\>$ in Hilbert space $\h_d\otimes \h_d$ with the Schmidt
decomposition
\begin{equation}\label{eq:entstate}
|\Phi\> =\sum_{i=0}^{d-1} \lambda_i |i\>|i\>
\end{equation}
where $\lambda_0\geq \dots\geq \lambda_{d-1}\geq 0$ and $\sum
\lambda_i^2=1$. In addition, there exists a noiseless
$d$-dimensional quantum channel by which Alice can send her particle
faithfully to Bob. The purpose of dense coding is to transmit from
Alice to Bob signals chosen from the set $\{0,1,\dots,d^2-1\}$. The
most general strategy is as follows. Alice encodes each possible
signal $r\in \{0,1,\dots,d^2-1\}$ into her particle by carrying out
a general quantum operation $\e_r$ on it, and then sends her
particle to Bob through the noiseless quantum channel. After
receiving Alice's particle, Bob performs a positive operator-valued
measure (POVM for short) $\{\Pi_s\}$ on the joint system. The
measurement result is used by Bob to make a guess on the signal
Alice sent to him. Here we do not specify the range of subscript $s$
in the POVM Bob performs, since as we will see in the following, it
varies in different dense coding schemes.

As already indicated in the Introduction, when the state in
Eq.(\ref{eq:entstate}) is partially entangled, i.e. $\lambda_0<1$,
perfect dense coding which can faithfully transmit $2\log_2 d$ bits
of classical information with certainty is impossible. If a certain
probability of error is permitted for Bob when guessing the signal
Alice sent, the task of dense coding can, however, be achieved. This
scheme is proposed in this paper and called approximate dense coding
with its aim to maximize the success transmission probability
\begin{equation}
    P_s = \sum_{r=0}^{d^2-1} p_r P(r|r).
\end{equation}
Here $p_r$ denotes the prior-probability of the occurrence of signal
$r$, and $P(s|r)$ denotes the probability of Bob retrieving signal
$s$ when $r$ was initially transmitted by Alice.

The probabilistic dense coding proposed in Refs.\cite{HLG00,PPA05}
and the unambiguous one in Ref.\cite{WCSG06}, on the other hand,
allow some probability with which the protocol fails with nothing
transmitted from Alice to Bob. Once it succeeds, however, the signal
Alice sent is recovered by Bob without error. That is, it is
required that
\begin{equation}\label{eq:constr}
    P(s|r)=0,\ \ \mbox{for all }s\neq r.
\end{equation}
The aim of this scheme is then maximize the conclusive probability
\begin{equation}\label{eq:maxpc}
    P_c = \sum_{r=0}^{d^2-1} p_r P(r|r)
\end{equation}
under the constraint of Eq.(\ref{eq:constr}).

In this paper, we adopt a different view of regarding unambiguous
dense coding as a protocol which can faithfully transmit a random
variable from the sender to the receiver. In other words, if the
signals Alice encoded in her particle were drawn upon a probability
distribution, after receiving Alice's particle, Bob should retrieve
the distribution perfectly with some probability. As a consequence,
by `unambiguous' we mean that not only any signal is transmitted
without error, but the post-probability of the signal occurring on
Bob's side, conditioning that conclusive results are obtained, is
the same as the prior-probability the signal was chosen on Alice's
side. To be specific, not only Eq.(\ref{eq:constr}) but also the
following constraint
\begin{equation}\label{eq:newc}
    P(r|con)=p_r
\end{equation}
is required, where $P(r|con)$ denotes the post-probability of the
outcome $r$ conditioning that conclusive results are obtained. The
aim of unambiguous dense coding in our sense is then maximize the
success probability in Eq.(\ref{eq:maxpc}) under the constraints of
Eqs.(\ref{eq:constr}) and (\ref{eq:newc}).

\section{Optimal dense coding: approximate case}

In this section, we consider dense coding in approximate case.
Suppose the operation $\e_r$ Alice carries out on her particle to
encode signal $r$ is represented by Kraus operators as follows:
\begin{equation}\label{eq:ei}
    \e_r(\rho)=\sum_k E_{rk}\rho E_{rk}^\dag,
\end{equation}
and the POVM Bob performs on the joint system has the form $\{\Pi_r:
r=0,1,\dots,d^2-1\}$. When the result $r$ is obtained, Bob declares
that the signal Alice sent is $r$. Suppose further that each $\Pi_r$
has the decomposition
\begin{equation}\label{eq:Pi}
    \Pi_r = \sum_t |\phi_{rt}\>\<\phi_{rt}|
\end{equation}
for some un-normalized states $|\phi_{rt}\>$. Here we omit the
ranges of the subscripts $k$ and $t$ since they are unimportant for
our discussion. Let the random variable $X$ upon which Alice chooses
the signals have the distribution $P(X=r)=p_r$. Then the success
probability of transmitting $X$ from Alice to Bob is
\begin{eqnarray}
    P_s &=& \sum_{r=0}^{d^2-1} p_r P(r|r)\nonumber\\
    &=&\sum_{r=0}^{d^2-1} p_r \tr
    (\Pi_r \sum_k (E_{rk}\otimes I)|\Phi\>\<\Phi| (E_{rk}^\dag\otimes I)).\label{eq:spro}
\end{eqnarray}

In what follows, we derive an upper bound on the average success
probability
\begin{equation}\label{eq:averagep}
\mathbb{E}P_s =\int P_s\text{d} \mathbf{p}
\end{equation}
of approximate dense coding over all possible random variables with
range $\{0,1,\dots,d^2-1\}$. Here the integral $\int \text{d}
\mathbf{p}$ over the space of $d^2$-dimensional probability
distributions is performed using the uniform measure. Techniques
used in the argument are mainly based on Ref.\cite{Ban00}.

To begin with, we write the vectors $|\phi_{rt}\>$ under the Schmidt
basis presented in Eq.(\ref{eq:entstate}) as
\begin{equation}\label{eq:phirt}
    |\phi_{rt}\>=\sum_{i=0}^{d-1} |\phi_{rt}^i\>|i\>,
\end{equation}
where again the vectors $|\phi_{rt}^i\>$ are not necessarily
normalized. From the completeness of the POVM operators $\Pi_i$, we
have
\begin{equation}\label{eq:sumtoI}
    I_{d}\otimes I_{d}=\sum_{r} \Pi_r =\sum_{r,t}\sum_{i,j=0}^{d-1} |\phi_{rt}^i\>\<\phi_{rt}^j|\otimes
    |i\>\<j|,
\end{equation}
and it follows that
\begin{equation}\label{eq:phicomplete}
   \sum_{r,t} |\phi_{rt}^i\>\<\phi_{rt}^j| = \delta_{ij}I_d
\end{equation}

Taking Eqs.(\ref{eq:entstate}), (\ref{eq:spro}), and
(\ref{eq:phirt}) into Eq.(\ref{eq:averagep}), we have
\begin{eqnarray}
    \mathbb{E}P_s&=& \sum_{r,k,t} (\int p_r\text{d} \mathbf{p}) \left|\<\phi_{rt}|E_{rk}\otimes I|\Phi\>\right|^2\nonumber\\
    &=&\frac{1}{d^2}\sum_{r,k,t} \left|\sum_i \lambda_i
    \<\phi_{rt}^i|E_{rk}|i\>\right|^2.\label{eq:spro2}
\end{eqnarray}
The last equality holds because $\int p_r\text{d} \mathbf{p}=d^{-2}$
for any $r=0,\dots,d^2-1$, which in turn is obvious from the
symmetry and the identity $\int \sum_r p_r\text{d} \mathbf{p}=1$.

We now recall the inequality
\begin{equation}\label{eq:equality}
    \sum_{\alpha=1}^N \left|\sum_{k=1}^M x_{k\alpha}\right|^2 \leq
    \left(\sum_{k=1}^M \sqrt{\sum_{\alpha=1}^N|x_{k\alpha}|^2}\right)^2
\end{equation}
from Ref.\cite{Ban00} which is simply the triangle inequality for
$M$ complex $N$-dimensional vectors ${\bf x}_k = (x_{k1},\ldots,
x_{kN})$ with the standard quadratic norm $||{\bf x}_k||^2 =
\sum_{\alpha=1}^{N} |x_{k\alpha}|^2$. So we proceed as
\begin{equation}\label{eq:spro3}
    \mathbb{E}P_s\leq\frac{1}{d^2} \left(\sum_i \lambda_i\sqrt{\sum_{r,k,t}
    \left|\<\phi_{rt}^i|E_{rk}|i\>\right|^2}\right)^2.
\end{equation}
The term under the square root in the above expression can be
further estimated by
\begin{eqnarray}
    &&\sum_{r,k,t}
    \left|\<\phi_{rt}^i|E_{rk}|i\>\right|^2\nonumber\\ &=& \sum_{r,t}
    \<\phi_{rt}^i|\phi_{rt}^i\>\sum_{k}\<i|E_{rk}^\dag\frac{|\phi_{rt}^i\>\<\phi_{rt}^i|}{\<\phi_{rt}^i|\phi_{rt}^i\>}E_{rk}|i\>\nonumber\\
    &\leq& \sum_{r,t}
    \<\phi_{rt}^i|\phi_{rt}^i\>\sum_{k}\<i|E_{rk}^\dag E_{rk}|i\>\nonumber\\
    &=& \sum_{r,t}
    \<\phi_{rt}^i|\phi_{rt}^i\>=d.
\end{eqnarray}
The last equality is due to Eq.(\ref{eq:phicomplete}). Notice that
we have implicitly assumed that $\<\phi_{rt}^i|\phi_{rt}^i\> \neq 0$
for any $r,t$ and $i$ in the above argument. There is no loss of
generality, however, since the above inequalities also hold when for
some $r,t$ and $i$, $\<\phi_{rt}^i|\phi_{rt}^i\>=0$. Thus finally we
arrive at our desired bound on the average success probability in
approximate dense coding which reads
\begin{equation}\label{eq:bound}
    \mathbb{E}P_s\leq \frac{1}{d}\left(\displaystyle\sum_{i=0}^{d-1}
    \lambda_i\right)^2.
\end{equation}

In the following, we construct explicitly a protocol which saturates
the bound presented in Eq.(\ref{eq:bound}). This protocol is in fact
the standard one for dense coding with higher dimensional maximally
entangled states. To simplify the notations, we introduce two
indexes $m$ and $n$ both taking values $0$ through $d-1$ to replace
the single index $r$ in the following argument. Let the operation
$\e_{mn}$ performed by Alice corresponding to signal $(m,n)$ be a
generalized Pauli operation $\sigma_{mn}$ such that
\begin{equation}\label{eq:sigma}
\sigma_{mn}=\sum_{k=0}^{d-1}e^{2\pi ikn/d}|k\oplus m\>\<k|
\end{equation}
where $\oplus$ denotes addition modulo $d$. Let the POVM carried out
by Bob be $\{\Pi_{mn}=|\phi_{mn}\>\<\phi_{mn}|,\ m,n=0,\dots,d-1\}$
where
\begin{equation}
|\phi_{mn}\>=\frac{1}{\sqrt{d}}\sum_{k=0}^{d-1}e^{2\pi
ikn/d}|k\oplus m\>|k\>.
\end{equation}
It is direct to check that $\sum_{mn} \Pi_{mn}=I_d\otimes I_d$. Here
in the above two equations, the basis $\{|k\>\}$ is the same as the
basis $\{|i\>\}$ presented in Eq.(\ref{eq:entstate}). We now
calculate the average success probability of this dense coding
protocol as
\begin{eqnarray}
    \mathbb{E}P_s&=&\frac{1}{d^2}\sum_{m,n=0}^{d-1}
    \left|\<\phi_{mn}|\sigma_{mn}|\Phi\>\right|^2\nonumber\\
    &=&
    \frac{1}{d^3}\sum_{m,n=0}^{d-1}\left|\sum_{k,k'=0}^{d-1}\lambda_{k'}
    e^{2\pi
i(k'-k)n/d}\<k\oplus m|k'\oplus m\>\right|^2\nonumber\\
    &=&\frac{1}{d^3}\sum_{m,n=0}^{d-1} \left(\sum_{k=0}^{d-1}\lambda_k
 \right)^2=\frac{1}{d} \left(\sum_{k=0}^{d-1}\lambda_k
 \right)^2.
\end{eqnarray}

\section{Optimal dense coding : unambiguous case}

We derive in the previous section the optimal strategy of
approximate dense coding with an arbitrary pure entangled state. In
this section, we consider the same problem in unambiguous dense
coding. As pointed out in Section II, what we are concerned with in
our notion of unambiguous dense coding is the transmission of random
variables, so the specific signal as well as the probability the
signal occurs must be recovered unambiguously on Bob's side. With
this criteria, the general strategy for Alice and Bob is as follows.
Alice carries out on her particle a quantum operation $\e_r$ to
encode the signal $r$, just as in approximate dense coding; while
Bob's POVM must have an additional element indicating the
inconclusive result. That is, the measurement should have the form
$\{\Pi_?,\Pi_r;r=0,1,\dots,d^2-1\}$. When the result corresponding
to $\Pi_?$ is obtained, the process fails, and nothing is
transmitted from Alice to Bob.

Let $p_r=P(X=r)$ be the probability of Alice choosing the operation
$\e_r$. Then the post-probability of the measurement outcome $r$
conditioning that conclusive results are obtained is
\begin{equation}\label{eq:postprior}
    P(r|con)=\frac{p_r P(con|r)}{\displaystyle\sum_r p_r
    P(con|r)}=\frac{p_r P(r|r)}{\displaystyle\sum_r p_r P(r|r)}.
\end{equation}
The second equality holds because $P(s|r)=\delta_{rs}P(r|r)$ which
in turn is due to the constraint that the dense coding scheme is
error-free.

By definition, unambiguous dense coding protocols must transmit
$any$ random variable from Alice to Bob unambiguously, so it is
required that $P(r|con)=p_r$ for any prior-probability distribution
$p_r$. Thus we deduce $P(r|r)=C$, $r=0,1,\dots,d^2-1$, for a
constant $C$ independent of $r$. That is, to achieve unambiguous
dense coding for any input random variable, the success probability
must be the same for each signal Alice wishes to send. We further
calculate the conclusive probability of the whole protocol as
\begin{equation}\label{eq:pc}
P_c =\sum_{r}p_r P(con|r)=\sum_{r}p_r P(r|r) = C,
\end{equation}
which is also independent of the distribution of the transmitted
random variable $X$. It has been proven in Ref.\cite{WCSG06} that in
the case of constant conditional success probability, it holds
$P_c\leq d\lambda_{d-1}^2$. So we finally have
\begin{equation}\label{eq:unbound}
\mathbb{E}P_c \leq d\lambda_{d-1}^2.
\end{equation}

It is worth noting that since the conclusive probability $P_c$ is
independent of the prior distribution of $X$, the bound presented in
Eq.(\ref{eq:unbound}) also applies if we take other quantities, e.g.
the maximum success probability over all possible prior-probability
distributions, as our criterions to judge the optimality of an
unambiguous dense coding protocol.

Somewhat surprisingly, the bound presented in Eq.(\ref{eq:unbound})
for unambiguous dense coding coincides with the bound proposed in
Ref.\cite{RDF03} for unambiguous teleportation. This can be regarded
as a new evidence for the close connection between dense coding and
teleportation.

To conclude this section, we construct an explicit protocol which
saturates the bound in Eq.(\ref{eq:unbound}). Notice that this bound
is just the maximal success probability of converting the partially
entangled state in Eq.(\ref{eq:entstate}) into a maximally entangled
state in the same Hilbert space using only local quantum operations
and classical communication \cite{LP01,Vid99}. A direct strategy for
Alice and Bob is first converting the shared entanglement into a
maximal one, and then utilizing this maximal entanglement to send
information perfectly using the standard protocol.

This strategy is, however, not the optimal one in the sense that
additional classical communication will be assumed in the process of
entanglement conversion. Fortunately, we can construct as follows a
direct protocol to saturate the bound without resorting to any
additional resource. Let the operation $\e_{mn}$ performed by Alice
corresponding to the signal $(m,n)$ be the general Pauli operation
$\sigma_{mn}$ defined in Eq.(\ref{eq:sigma}), just as in the optimal
approximate protocol. The POVM carried out by Bob, however, has the
following form
\begin{equation}\label{eq:unmeasure}
    \Pi_{mn}=\frac{\lambda_{d-1}^2}{d}|\phi_{mn}\>\<\phi_{mn}|
\end{equation}
for  $m,n=0,\dots,d-1$, and
\begin{equation}\label{eq:unincon}
\Pi_? = I_{d}\otimes I_d-\sum_{m,n=0}^{d-1} \Pi_{mn}
\end{equation}
where the un-normalized state
\begin{equation}
|\phi_{mn}\>=\sum_{k=0}^{d-1}\lambda_k^{-1} e^{2\pi ikn/d}|k\oplus
m\>|k\>.
\end{equation}
Without loss of generality, we assume that $\lambda_{d-1}>0$ (so
$\lambda_k>0$ for any $0\leq k\leq d-1$) because otherwise the bound
is equal to 0, and it can be saturated trivially.

We now prove that the set of measurement operators in
Eqs.(\ref{eq:unmeasure}) and (\ref{eq:unincon}) indeed constitute a
POVM. This can be validated from the calculation
\begin{eqnarray}
    \sum_{m,n=0}^{d-1} \Pi_{mn} &=&  \lambda_{d-1}^2
    \sum_{m,k} \lambda_{k}^{-2}
    |k\oplus m\>\<k\oplus m|\otimes|k\>\<k|\nonumber\\
    &=& I_d\otimes \sum_k
    \left(\frac{\lambda_{d-1}}{\lambda_k}\right)^2|k\>\<k|\nonumber \\
    &\leq & I_{d}\otimes I_d.
\end{eqnarray}

Furthermore, for any signals $(m,n)$ and $(m',n')$, we have
\begin{eqnarray}
    &&\tr(\Pi_{mn}\e_{m'n'}(|\Phi\>\<\Phi|))=\frac{\lambda_{d-1}^2}{d}|\<\phi_{mn}|\sigma_{m'n'}|\Phi\>|^2\nonumber\\
    &&=\frac{\lambda_{d-1}^2}{d}\left|\sum_{k,k'}\lambda_{k'}\lambda_{k}^{-1}e^{2\pi i(kn'-kn)/d}\<k\oplus m|k\oplus m\>\<k|k'\>\right|^2\nonumber\\
    &&=d\lambda_{d-1}^2\delta_{mm'}\delta_{nn'}.
\end{eqnarray}
The probability of conclusively transmitting the signal $(m,n)$ is
then $P[(m,n)|(m,n)] = d\lambda_{d-1}^2,$ which is independent of
the specific signal $(m,n)$ and the prior-distribution. From these
facts, it is easy to show that this protocol can indeed
unambiguously transmit any random variable from Alice to Bob, and
also, the bound presented in Eq.(\ref{eq:unbound}) for unambiguous
dense coding is reached.
\section{Conclusion}
In this paper, we present optimal dense coding strategies for
approximate and unambiguous cases when partial entanglement between
the sender and the receiver is provided. These strategies are
optimal in the sense that the average success probability (in
approximate case) or the average probability of conclusive results
(in unambiguous case) is maximized. Notice that the optimal average
success probability for approximate dense coding given in
Eq.(\ref{eq:bound}) depends on the sum of all the Schmidt
coefficients $\sum_{i=0}^{d-1}\lambda_i$, while in unambiguous dense
coding, the optimal average conclusive probability presented in
Eq.(\ref{eq:unbound}) depends only on the least Schmidt coefficient
$\lambda_{d-1}$. These results give new evidences to the
correspondence between dense coding and teleportation since the same
dependency can be found in approximate teleportation \cite{Ban00}
and unambiguous teleportation \cite{RDF03}.

\section*{Acknowledgement}
The authors thank the colleagues in the Quantum Computation and
Quantum Information Research Group for useful discussion. This work
was partly supported by the Natural Science Foundation of China
(Grant Nos. 60503001, 60321002, and 60305005), and by Tsinghua Basic
Research Foundation (Grant No. 052220204). R. Duan acknowledges the
financial support of Tsinghua University (Grant No. 052420003).

\bibliography{densecoding}

\end{document}